\documentclass[12pt,preprint]{aastex}

\title{OB Stars in the Solar Neighborhood II: Kinematics}

\author{F. Elias}
\affil{Facultad de F\'{\i}sica. Departamento de F\'{\i}sica
At\'omica, Molecular y Nuclear. Universidad de Sevilla, Apartado
1065, Sevilla, Spain.}

\author{E.J. Alfaro}
\affil{Instituto de Astrof\'{\i}sica de Andaluc\'{\i}a, CSIC,
Apartado 3004 Granada, Spain.}

\author{J. Cabrera-Ca\~no}
\affil{Facultad de F\'{\i}sica. Departamento de F\'{\i}sica
At\'omica, Molecular y Nuclear. Universidad de Sevilla, Apartado
1065, Sevilla, Spain.} \affil{Instituto de Astrof\'{\i}sica de
Andaluc\'{\i}a, CSIC, Apartado 3004 Granada, Spain.}

\begin{document}

\begin{abstract}
  Using the spatial classification method and the structural
  parameters estimated for the Gould Belt (GB) and the local Galactic
  disk (LGD) from a previous paper, we have evaluated spatial
  membership probabilities for a sample of O and B stars from the {\it
  Hipparcos\/} catalogue \citep{ESA97} with available proper motions
  and radial velocity data. Thus being able to study the space
  velocity fields of both systems, we conclude that the GB and the LGD
  present different statistical distributions, both in the velocity
  space as well as in the phase space. In the light of the existence
  of both systems we analyze different kinematic aspects in the
  vicinity of the Sun, and we find the GB to be responsible for the
  highly negative vertex deviation found for the OB stars in the solar
  neighborhood. We also find that the GB sensibly alters the estimation of Oort's
  constants in the Galactic disk; thus, in order to establish comparisons with other kinematic
  studies based in older stellar populations, a careful pruning of the GB members must be
  performed. Further analysis of the GB velocity field and the
  moving groups that contribute to it suggest the possibility that the GB can be
  the result of a chance superposition of moving groups. We propose
  for future investigations the dynamical analysis of these moving
  groups in order to probe the origin of the GB.

\end{abstract}

\keywords{(Galaxy:) solar neighborhood --- stars: early-type ---
stars: kinematics}

\section{Introduction}
Since the discovery of the GB as a system of bright stars forming an
angle of about $20\degr$ with the Galactic disk \citep{Gou79,Her47},
many efforts have been devoted to unravel its complex structure of
stars and the associated interstellar medium. When studies of its
kinematics began to appear on the second half of the past century,
the peculiarities of its behavior made the global picture of the GB
even more puzzling.

The most striking discovery was the expansion motion of the stellar
component \citep{Bon64,Les68,Fri75,Fro77,Wes85,Com94}, which cannot
be satisfactorily explained by a single explosive event
\citep{Les68,Mor99}, making it difficult to trace back the origin of
the GB by reversing this movement. Models considering Galactic
density waves generated by perturbations from the spiral arms cannot
account for this expansion either \citep{Wes85,Com91}. Also, the
shell of gas associated with the GB present evidences of an
expansion movement, according to the works of \citet{Lin67},
\citet{Ola82}, \citet{Elm82} and \citet{Mor99}.

Because the age of the GB (between 20 and 90 Myr, see \citet{Tor00}
for a detailed discussion) is a considerable fraction of the period
of vertical oscillation of the stars over the Galactic plane under
the influence of the gravitational potential of the Galaxy, the
spatial coherence of the GB must be corresponded by a kinematical
coherence that prevents the dissolution of the structure among the
Galactic plane. \citet{Com99} and \citet{Per03} find a global
oscillation movement of the GB around an axis different from the
spatial line of nodes where the GB and the Galactic plane cross each
other.

All this leads \citet{Elm00} to include the GB as part of the
hierarchy of stellar complexes within the Milky Way, as a second
level structure subordinated to the local arm (the Orion-Cygnus
spur). Thus, the GB would be the star formation complex closest to
us, so the study of its global properties will contribute to shed
light on the possible origin and evolution of these complexes.

In a previous article (Elias et al. 2006, from now on referred to as
Paper I) we had developed a three-dimensional spatial classification
method to separate the GB stars from the LGD stars. Having obtained
the structural parameters of both systems exclusively by spatial
considerations, we now shall classify a sample of {\it Hipparcos\/}
\citep{ESA97} OB stars with space velocities, in order to compare
the kinematics of both structures.

Thus, in Section 2 we first build a sample of O-B6 {\it Hipparcos\/}
stars with proper motions and radial velocity data, that we then
analyze in order to study the kinematic properties of the young
Galactic disk. We begin (Section 3) by the identification of the
moving groups in the velocity field, and then (Section 3.1) we
separate the GB from the LGD stars, which allows us to decide about
the membership to either system of the detected moving groups. The
next step to enhance our analysis of the young Disk velocity field
is the elimination of the systematic effects on the velocities by
the solar motion and the Galactic differential rotation. We obtain
the residual velocities, the analysis of which also yields a
different kinematic behavior of the GB and the LGD (Section 4). The
study of the velocity ellipsoids confirms such difference (Section
4.1). The effect of the GB on the determination of the Oort
constants is addressed in Section 5. Finally, some conclusions are
exposed in Section 6.

\section{Star sample}
We select a sample of 1156 stars of spectral types O-B6 and
luminosity classes III, IV and V from the {\it Hipparcos\/}
catalogue \citep{ESA97}. Photometric data from the \citet{Hau98}
catalogue, as well as radial velocity data from the \citet{Bar00}
and \citet{Gre99} catalogues have been added when available. Thus,
the compilation includes:
\begin{itemize}
\item HIP, {\it Hipparcos\/} identifier number
\item Spectral type
\item $V$, Johnson visual magnitude
\item Trigonometric parallax (mas)
\item Standard error in trigonometric parallax (mas)
\item Right ascension for the epoch J1991.5 in ICRS (degrees)
\item Declination for the epoch J1991.5 in ICRS (degrees)
\item Proper motion in right ascension, $\mu_{\alpha} cos{\delta}$
(mas yr$^{-1}$)
\item Standard error in proper motion in right ascension (mas
yr$^{-1}$)
\item Proper motion in declination, $\mu_{\delta}$ (mas yr$^{-1}$)
\item Standard error in proper motion in declination (mas yr$^{-1}$)
\item Radial velocity (km s$^{-1}$) from \citet{Bar00}
\item Quality of radial velocity from \citet{Bar00}
\item Radial velocity (km s$^{-1}$) from \citet{Gre99}
\item Error in radial velocity (km s$^{-1}$) from \citet{Gre99}
\item $uvby\beta$ photometry data from \citet{Hau98}
\end{itemize}

For the distance estimation we have used {\it Hipparcos\/}
trigonometric parallaxes only if the relative error is lower or
equal to a $10\%$. Otherwise, $uvby\beta$ Str\"omgren photometry has
been used to estimate the distance through the \citet{Bal87}
$M_{V}(\beta)$ calibration. If no data were available,
spectro-photometric distances from the apparent visual magnitude $V$
and the \citet{Sch82} calibration for the spectral types have been
chosen. We have compared the three different distance estimations,
looking for any possible systematic biases among them. With this
purpose, we have selected 950 stars from the initial catalog with
Str\"omgren photometry data.

Several authors \citep{Are99,Mai01,Mai05,Sch04}, studying the
distance calibration comparisons with the distances obtained from
trigonometric parallaxes, have dealt with the problem of analyzing
the biases that the sample selection effects introduce. The
estimation of these biases is very complex because it depends, among
other variables, on the spatial distribution of the sample.
\citet{Mai05} demonstrates that the real distance probability
distribution for individual stars will always be ill-behaved when
that distance tends to infinite and a constant underlying spatial
distribution has been assumed for the sample. This idea had already
been suggested by \citet{Are99}, who propose that -in order to avoid
any truncation biases- for comparison purposes it should be used a
sample not selected by any limit in the relative error of the
trigonometric parallax, including also the negative parallaxes. We
use this methodology for the comparison between trigonometric,
photometric and spectroscopic parallaxes, using the complete sample
with spectral types up to B6.

The result of the analysis demonstrates that the Str\"omgren
photometric parallaxes ($\pi_{Str}$) are very similar to the {\it
Hipparcos\/} trigonometric parallaxes -as \citet{Kal98} had already
demonstrated-, but the former present, in comparison with the
spectroscopic parallaxes ($\pi_{SK}$, derived from the Schmidt-Kaler
calibration), a functional relationship in the form of
$\pi_{Str}/\pi_{SK} \sim 1.21$. Nevertheless, a recent study for the
O stars \citep{Mai05} demonstrates that the {\it Hipparcos\/}
trigonometric parallaxes and the spectroscopic parallaxes are very
similar for this spectral type. Where does this difference that we
find in their values come from? We have to consider that our
spectroscopic calibration is based on three steps: 1.- Spectral
classification; 2.- Calibration of the absolute magnitude for each
spectral type; 3.- Evaluation of the reddening from the intrinsic
color values.

The spectral classification of the {\it Hipparcos\/} catalog comes
from different sources and thus is far from being uniform. On the
other hand, the catalog of O stars \citep{Mai04} used by
\citet{Mai05} for the comparison of the parallaxes was derived from
a group of homogeneous and precise stellar spectra. Thus, it must
not be inferred from our result that the Schmidt-Kaler calibration
has systematic biases, but that the spectroscopic parallaxes of the
O-B6 stars from the {\it Hipparcos\/} catalog -according to the
spectral classification within the catalog- show a systematic error
when we compare them with the Str\"omgren photometric parallaxes
(which, as we said, bear no significative difference with the {\it
Hipparcos\/} trigonometric parallaxes).

In this work, we do not intend to perform an exhaustive study of the
problems that the diverse methodologies of obtaining distance
calibrations arise. We have just unified for working reason the
different distance calibrations used for our sample. For that, we
have tied the spectroscopic parallaxes to those derived from
Str\"omgren photometry, applying a correction of 21\% to the former.
We want to stress that the comparison between the different distance
calibrations has been made strictly through their respective
parallaxes.

We have also chosen the \citet{Bar00} radial velocity and quality
data when available; otherwise, radial velocities and errors from
\citet{Gre99} have been used. Finally, a distance limit of 1 kpc has
been imposed, thus reducing the sample to 881 stars.

While the {\it Hipparcos\/} catalogue is complete down to
$V\sim7.9$, and to $V\sim7.5$ for O-B6 stars, in Figure 1 we can see
through a histogram that our sample is complete only down to a
magnitude of $V\sim6.5$. This is caused by the necessity of having
radial velocity data available for the stars in our sample, as it is
also observed in \citet{Dav05}; in that work, the completeness of
the sample of O and B stars from the {\it Hipparcos\/} catalogue
falls down from $V\sim7.9$ to $V\sim6$ when the stars without radial
velocity data are removed.

\section{Identification of moving groups in the sample and their membership}

We have calculated the space velocities from the proper motions and
the radial velocities \citep{Joh87} for the stars in our sample.
Their density field is represented in the three contour plots of
Figure 2. We appreciate in this image that the velocity field is
dominated by several maxima which may correspond to associations of
stars (not necessarily bound) with a small velocity dispersion,
known as moving groups (e.g. Proctor 1869; Eggen 1963).

The first and most prominent maximum, located around $(U, V, W) =
(-6.5, -19, -7)$ km s$^{-1}$, is certainly associated with the
Pleiades moving group. The exact situation of the peak may differ
slightly from the estimation given by other authors, but we must
consider that the moving group always appears as a maximum in the
velocity space with a certain width. For instance, \citet{Che97},
working with a sample of B, A and F stars from the Hipparcos Input
Catalogue \citep{Tur92} and $uvby\beta$ photometry, find that the
maximum is located at $(U, V, W) = (-10, -19, -8.1)$ km s$^{-1}$,
the standard deviation for the three components being, respectively,
$(7.9, 8.6, 5.8)$ km s$^{-1}$.

The second most prominent peak in Figure 2 is located around $(U, V,
W) = (-17, -11, -5)$ km s$^{-1}$. We have identified it as the
moving group related to the supercluster IC 2391. Its position in
velocity space is estimated by \citet{Che97} at $(U, V, W) = (-15.9,
-13.1, -4.5)$ km s$^{-1}$, the standard deviation being
$(\sigma_{U}, \sigma_{V}, \sigma_{W}) = (4.1, 6.2, 3.0)$ km
s$^{-1}$.

The third maximum, centered around $(U, V) = (-11, -8.5)$ km
s$^{-1}$, is more diffuse and difficult to identify. We have decided
to follow the criteria of \citet{Asi99}, who -based on the studies
by \citet{Com92}- rule out the possibility of linking this region in
the velocity space with the Coma Berenices cluster, in favor of a
probable bond with the Cassiopeia-Taurus association, of $(U, V) =
(-9.9, -6.1)$ km s$^{-1}$. This association has an age between 50
Myr -as the probable expansion age estimated by \citet{Bla56} from a
sample of stars of spectral type B5 or earlier- and $90\pm10$ Myr,
as determined from the Lithium depletion boundary by \citet{Sta99}.
In the detailed study of the nearby OB associations by \citet{Zeu99}
the authors find a physical relation between the Cas-Tau group and
the $\alpha$ Persei cluster. The main sequence turn-off age for this
cluster is about 50 Myr \citep{Mey93}, so the age of both Cas-Tau
and $\alpha$ Persei could be the same (e.g. Brown 2002). Note that
in Paper I we found that $\alpha$ Persei is located well within the
spatial boundaries of the GB.

\subsection{Classification of the sample}

In order to observe how these moving groups contribute to the two
systems in study, we classify the sample and determine the
membership of the stars to either the GB or the LGD. In Paper I we
obtained several estimations of the parameters that characterize
both systems in our model. We shall work with the solution for the
full O-B6 sample of Paper I with an exponential model for the
stellar distribution in the direction perpendicular to the mean
planes.

Thus, using that estimation as the true value of the GB and the LGD
mean planes, we simply apply our separation algorithm to our current
star sample with kinematic data. No iteration in order to
re-evaluate the mean planes is performed, we just assign membership
probabilities to the stars according to the planes estimated in
Paper I. Also, spatial outliers are eliminated according to the
procedure explained in that paper; the remaining sample has 776
stars. We obtain a separation between the GB and the LGD based
exclusively in the spatial position of these stars. Yet we can see
in Figure 3 how a difference in their velocity fields is obtained as
a result.

The most striking difference is that in the $UV$ projection (top
panels of Figure 3) the three moving groups that we had found in the
full sample (top panel of Figure 2) now distinctly belong either to
the GB or the LGD. The two maxima associated with the Pleiades and
IC 2391 appear only in the GB velocity field (Figure 3, top left
panel), while the Cassiopeia-Taurus peak remains only visible in the
LGD field (Figure 3, top right panel). This is not surprising, if we
consider that the Pleiades moving group is spatially related to the
Sco-Cen association, which is one of the main components of the GB
\citep{Mor99}. Also, we know that IC 2391 is a young cluster, its
age being about 30 Myr \citep{Sta97}. In a recent study of the
tangential velocities, \citet{Pis06} conclude that its kinematic
probability of belonging to the GB is a $73\%$. Note that we have
arrived to a similar conclusion by a process based solely on the
spatial position of the stars, and thus independent of the result in
the cited paper.

We must also note that a late-type population of young stars has
been associated to both the Pleiades and IC 2391 by \citet{Mon01}.
In that paper, these moving groups are described as centered around
$(U, V, W) = (-11.6, -21, -11.4)$ and $(U, V, W) = (-20.6, -15.7,
-9.1)$ km s$^{-1}$, respectively, with a dispersion of about 8 km
s$^{-1}$ around the central positions. Not surprisingly, a late-type
component of stars of about 30-80 Myr of age had already been
associated to the GB disk structure by \citet{Gui98} studying the
X-ray sources in the ROSAT All-Sky Survey.

Finally, we also observe that in the $UW$ projection (Figure 3,
middle right panel) it clearly rises a new maximum that has a
correspondence with a small protuberance around $U = 11$ km s$^{-1}$
in the $UV$ projection (Figure 3, top right panel). It was also
present, although weak, in the velocity density field of the full
sample (Figure 2, top and middle panels). We haven't found an exact
correspondence to this possible moving group among the structures in
the solar neighborhood, but the positive value of the $U$ component
makes us think that it may be related to the Sirius supercluster
(eg. Eggen 1996, Dehnen 1998, Asiain et al. 1999).

\section{Analysis of the residual velocities}
If we want to improve our study of the kinematic differences between
the GB and the LGD, we must work with residual velocities. This way,
the systematic effects of the Galactic kinematics will not interfere
with our comparison. Thus, we now correct the space velocities of
the stars in our sample from solar motion (using the classical
estimation of \citet{Del65}, ($U_{\odot}$, $V_{\odot}$, $W_{\odot}$)
= (9, 12, 7) km s$^{-1}$) and from Galactic differential rotation
-using the values given by \citet{Ker86}.

In order to refine the analysis of the velocity distributions, we
must also eliminate outliers of kinematic nature that may be present
in the sample. That is, we eliminate the stars located in the
regions of a very low density in the velocity space for our sample
under study. We achieve this by running the OUTKER algorithm
\citep{Cab85}, which reduces the sample to a final number of 752
stars. We then separate again the sample into GB and LGD members.Now
we want to compare both distributions in a N-dimensional space,
where N=6 (phase space) or N=3 (velocity space). In order to achieve
this, we employ a multidimensional, non-parametric two-samples test:
the Cramer test \citep{Bah04}. The test statistic is the difference
of the sum of all the Euclidean interpoint distances between the
random variables from the two different samples:

\begin{equation}
T = \frac{mn}{m+n} \left(\frac{2}{mn} \sum_{i,j}^{m,n}
\phi(||\textbf{\emph{X}}_i-\textbf{\emph{Y}}_j||^2) - \frac{1}{m^2}
\sum_{i,j=1}^{m} \phi(||\textbf{\emph{X}}_i-\textbf{\emph{X}}_j||^2)
- \frac{1}{n^2} \sum_{i,j=1}^{n}
\phi(||\textbf{\emph{Y}}_i-\textbf{\emph{Y}}_j||^2) \right)
\end{equation}

where $\textbf{\emph{X}}_i$ and $\textbf{\emph{Y}}_i$ are the point
vectors of each sample member, and m,n are the respective size of
the samples. $\phi$ is a kernel function; for this particular case
we have used the Cramer kernel implemented by \citet{Fra04} in the R
statistical environment \citep{R05}.

Thus, an analysis of the three-dimensional velocity space, $(U, V,
W)$, yields that the GB and the LGD distributions are different with
a 99\% of confidence. Similarly, the Cramer test for the six
dimensions of the phase space, $(X, Y, Z, U, V, W)$, rejects the
possibility that the GB and the LGD distributions are the same with
a confidence of a 99\%.

Thus we confirm that the GB and the LGD are two different stellar
systems in the sense that they show a clear statistical separation
between their distributions in the phase space. It is unavoidable to
take into consideration the contribution of the GB when the young
Disk is under study, because, as we have demonstrated, the
velocities of the OB stars in the solar neighborhood are not
statistically independent of their spatial positions.

\subsection{Velocity ellipsoids}
A more intuitive visualization of the differences between the GB and
the LGD residual velocity distributions are the contour density
plots of the three different velocity planes (see Figure 4). Their
disparate shapes and orientations already show that the velocity
ellipsoids clearly reflect the distinct kinematic behavior of both
stellar systems.

We have estimated the main geometrical parameters of the velocity
ellipsoids for the whole sample and for the GB and the LGD members
separately. The results are displayed on table~\ref{tbl-1}. Two main
results arise from this analysis:
\begin{enumerate}
\item  The vertex deviation for the GB is highly negative, while
that for the LGD is positive.

\item There is the suggestion that the third axis of the velocity ellipsoid for the GB is tilted in respect to the Galactic plane.
\end{enumerate}

The estimation of the vertex deviation in the solar neighborhood
from different star samples has produced different values depending
on the nature of the sample and on the kinematic variables used in
the calculation (see \citet{Mor99} for a compilation of previous
results). In brief, the general conclusion has been that the vertex
deviation becomes more negative as the star sample is younger. In
fact, one of the classic estimates based on space velocities for OB
stars \citep{Fil57} yielded a value close to $l_{v} = -50\degr$ for
the Galactic disk. For years, this result has remained a puzzling
issue that has been given several and varied explanations. Most of
them can be classified into two classical types: nature or nurture,
we could say. Some authors claim that these young stars that formed
from a molecular cloud show the same kinematics as the parent cloud
at the time of the star formation. This way, the initial velocity
and later expansion define the velocity ellipsoid of the present
stellar system. Other authors, though, defend that the effect of
different singular and punctual events (such as the passing through
a spiral arm) could also be the cause of the peculiar velocity
ellipsoid observed in the young stellar component.

What we conclude from our analysis is that the classic problem of
the negative vertex deviation for young stars in the solar
neighborhood is a consequence of the presence of the GB. If we
eliminate the stars that belong to the GB, the remaining sample of
only LGD stars present a positive vertex deviation ($l_{v}^{LGD} =
18\degr$).

\citet{Mor99}, working with a sample of dwarf O-B5.5 stars members
of the GB, found that the negative vertex deviation ($l_{v} \sim
-64\degr$) was caused by the Pleiades moving group. Once removed
this group, they obtained a positive vertex deviation ($l_{v} =
22\degr$) for the remaining stars in the GB. We now demonstrate the
the OB stars of the LGD also have a positive vertex deviation,
similar to that found by \citet{Mor99} for the GB stars after
eliminating the ones belonging to the Pleiades moving group. So when
we work with samples (either of GB or LGD stars) in which a single
moving group clearly dominates over the others the vertex deviation
is positive and close to $l_{v} \sim 20\degr$, which is the value
expected from the dynamic equations of the system for this age
group. The negative vertex deviation of the GB seems to be
originated by the relative position of the centroids of two moving
groups rather than by the distribution of the residual velocities as
a whole. We must note that the analysis of the young stars
-specially those belonging to the GB- based on the Schwarzschild
distribution from which the velocity ellipsoid is tailored, is not
the best suited to describe the reality of the system. The velocity
field is dominated, as we have seen, by the presence of moving
groups, and thus it is far from the hypothesis of a homogeneous and
stationary system. The problem thus lies in that the classic
analysis of the velocity ellipsoid is applied to a set of moving
groups. Already \citet{Mih81} pointed out that the cause of the
vertex deviation was the existence of moving groups in the Galactic
velocity field. But even though the velocity ellipsoid does not
strictly correspond to a physic reality in our case, that doesn't
invalidate our result, that is, the fundamental contribution of the
GB to the negative vertex deviation for the O and B stars.

Thus, the different values of the vertex deviation found in the
literature can be explained according to the different proportions
of GB stars present in the respective samples. This translates the
question about the origin of the negative vertex deviation to the
investigation of the origin of the moving groups. Although the
latter is out of the scope of this paper, we want to stress that the
more we study in detail the nature of the GB, the more we find
indications that we must probe both its nature and origin as a set
of moving groups.

\begin{deluxetable}{crrr}
\tablecolumns{4} \tablewidth{0pt} \tabletypesize{\scriptsize}
\tablecaption{Velocity ellipsoids\label{tbl-1}}
\tablehead{\colhead{} & \colhead{$\sigma$}
& \colhead{$l$} & \colhead{$b$} \\
\colhead{$Axis$} & \colhead{(km s$^{-1}$)} & \colhead{$(\degr)$} &
\colhead{$(\degr)$}} \startdata
\sidehead{Full sample:}
           $U'$ & $10.7 \pm 0.4$ & $-5 \pm 29$ &  $6 \pm 4$\\
           $V'$ & $10.2 \pm 0.4$ & $85 \pm 29$ & $-2 \pm 4$\\
           $W'$ & $6.9 \pm 0.6$ &  $153 \pm 115$ & $84 \pm 3$\\
\sidehead{Gould Belt:}
           $U'$ & $9.8 \pm 0.3$ &  $-47 \pm 22$ &  $-10 \pm 8$\\
           $V'$ & $8.9 \pm 0.3$ &  $43 \pm 22$  &    $1 \pm 12$\\
           $W'$ & $7.1 \pm 0.8$ &  $141 \pm 92$ &  $80 \pm 12$\\
\sidehead{Local Galactic disk:}
           $U'$ & $12.2 \pm 0.6$ & $18 \pm 11$ & $-2 \pm 4$\\
           $V'$ & $10.2 \pm 0.5$ & $-72 \pm 11$ & $0 \pm 5$\\
           $W'$ & $7 \pm 1$ &  $27 \pm 66$ & $88 \pm 5$\\
\enddata
\tablecomments{$U'$, $V'$ and $W'$ represent the principal axes of
the ellipsoids, which deviate from the $(U,V,W)$ reference frame.
The errors were estimated by bootstrap.}
\end{deluxetable}

Another striking result is the inclination of the GB's velocity
ellipsoid of about $10\degr \pm 12\degr$ with respect to the $UV$
plane, although this value does not have a great statistical
significance. While the third axis ($W'$) of the LGD ellipsoid
merely shows an inclination of $2\degr$, the GB's ellipsoid
inclination (10\degr) reminds that of the spatial system with
respect to the Galactic plane ($i_{GB} \sim 14\degr$ as we found in
Paper I). Some authors have pointed out that the GB could be
oscillating as a whole around the Galactic plane
\citep{Com99,Per03}. But our result does not discard a different
interpretation in terms of a juxtaposition of moving groups: this
links back to the idea that a descriptive parameter of the GB -such
as this inclination of the minor axis of the ellipsoid- can be
interpreted in terms of the relative position (in the velocity
space) of two moving groups.

\section{Estimation of the Oort constants}
Although our sample hasn't been ideally compiled with the intention
of calculating the Oort constants, it is worth performing a basic
estimation to obtain information about the GB's contribution to
their value. Thus, within the axisymmetric approximation of Oort's
model in first order, and including a $K$ expansion term, we solve
the condition equations \citep{Sma68,Clu72,Fro77} for the radial and
the tangential velocities:

\begin{equation}
v_{r} = u_{0} \cos l \cos b + v_{0} \sin l \cos b + w_{0} \sin b + A
r \sin 2l \cos^{2} b + K
\end{equation}
\begin{mathletters}
\begin{eqnarray}
v_{l} = -u_{0} \sin l + v_{0} \cos l + A r \cos 2l \cos b + B r \cos b\\
v_{b} = -u_{0} \cos l \sin b - v_{0} \sin l \sin b + w_{0} \cos b -
A r \sin 2l \cos b \sin b
\end{eqnarray}
\end{mathletters}

where $v_{r}$ is the radial velocity; $v_{l} = 4.74057$ $\mu_{l} r$
and $v_{b} = 4.74057$ $\mu_{b} r$ are the tangential velocities in
Galactic longitude ($l$) and latitude ($b$), being $\mu_{l}$ and
$\mu_{b}$ the respective proper motions and $r$ the heliocentric
distance; and $(u_{0},v_{0},w_{0}) =
-(U_{\odot},V_{\odot},W_{\odot})$, the reflex of the solar motion.

In principle, it would only make sense to estimate Oort's constants
for the LGD, eliminating the GB members from the sample. As we have
seen, from the study of the moving groups and the velocity
ellipsoid, the velocity distribution of the GB clearly deviates from
the axisymmetric hypothesis; hence, the Oort constants wouldn't
correspond to a physical reality in this case. Yet we have solved
the equations in order to establish a comparison and thus observe
the effects that the presence of the GB introduces in the kinematics
of the LGD.

The solutions for the proper motions and for the radial velocity
alone are listed on table~\ref{tbl-2}. The solar motion for the LGD
proper motions solution is in good agreement with the IAU standard,
$(U_{\odot},V_{\odot},W_{\odot})$ = (9, 12, 7) km s$^{-1}$
\citep{Ker86}. But the Oort constants differ noticeably from the IAU
recommended values of $A$ = 14.4 $\pm$ 1.2 km s$^{-1}$ kpc$^{-1}$
and $B$ = -12.0 $\pm$ 2.8 km s$^{-1}$ kpc$^{-1}$, and instead we
have a flat rotation curve with $A$ = $-B$ = 16 km s$^{-1}$
kpc$^{-1}$. Yet the value provided by \citet{Ker86} is a mean over
the results obtained by several authors, so it is more revealing to
compare our estimation with that of a study similar to ours.

\begin{deluxetable}{lrrrccr}
\tablecolumns{7} \tablewidth{0pt} \tabletypesize{\scriptsize}
\tablecaption{Oort constants\label{tbl-2}} \tablehead{\colhead{} &
\colhead{$U_{\odot}$} & \colhead{$V_{\odot}$} &
\colhead{$W_{\odot}$} & \colhead{$A$} & \colhead{$B$} & \colhead{$K$} \\
\colhead{$Sample$} & \colhead{(km s$^{-1}$)} & \colhead{(km
s$^{-1}$)} & \colhead{(km s$^{-1}$)} & \colhead{(km s$^{-1}$
kpc$^{-1}$)} & \colhead{(km s$^{-1}$ kpc$^{-1}$)} & \colhead{(km
s$^{-1}$)}}

\startdata

\sidehead{Solution from proper motions:}
$FS$ & $9.8 \pm 0.4$ & $13.0 \pm 0.6$ & $6.6 \pm 0.3$ &  $14 \pm 1$ & $-18 \pm 1$ & -\\
$GB$ & $9.9 \pm 0.5$ & $13.0 \pm 0.6$ & $6.7 \pm 0.3$ &   $11 \pm 2$ & $-20 \pm 1$ & -\\
$LGD$&  $9.4 \pm 0.8$ & $12.6 \pm 0.9$ & $6.3 \pm 0.3$ &  $16 \pm 2$ & $-16 \pm 1$ & -\\
\sidehead{Solution from radial velocity:}
$FS$ & $8.7 \pm 0.8$  & $15.0 \pm 0.6$ & $11 \pm 2$ & $13 \pm 2$ & - & $-0.6 \pm 0.4$\\
$GB$ &    $10 \pm 2$  &     $15 \pm 1$ &  $8 \pm 4$ &  $9 \pm 2$ & - &  $0.4 \pm 0.6$\\
$LGD$&     $9 \pm 1$  &     $14 \pm 1$ & $8 \pm 5$ & $16 \pm 3$ & - &     $-2 \pm 1$\\
\enddata
\tablecomments{FS is the Full Sample}
\end{deluxetable}

\citet{Fro77}, working with a sample of O-B5 stars, separate the GB
from the LGD, and considering a fixed value of $A$ = 15 km s$^{-1}$
kpc$^{-1}$, they estimate the value of $B$ for both systems, as well
as for the whole unclassified sample. For the latter they find a
flat rotation curve of $A$ = $-B$ = 15 km s$^{-1}$ kpc$^{-1}$; for
the LGD, $B$ = -12 km s$^{-1}$ kpc$^{-1}$; and for the GB, $B$ = -19
km s$^{-1}$ kpc$^{-1}$, a value similar to our estimation. But
regardless of the numerical results, we can conclude too that the
global kinematics of the GB differs greatly from the kinematics of
the LGD, and that the presence of this stellar system affects the
estimation of the parameters describing the kinematics of the solar
neighborhood.

On the other hand, the $K$ term is generally inferior in absolute
value to the estimation by \citet{Fro77}. We have checked that when
solving the equations without this term, the results for the Oort
constants do not change significatively. Thus we do not consider it
wise to conclude anything about the possible expansion movements of
the system from these results.

The detailed study of the local irregularities in the kinematics of
the young stars by \citet{Tor00}, from a sample of O and B {\it
Hipparcos\/} stars, reveals that for heliocentric distances lower
than 600 pc and age groups under 90 Myr (i.e., for a sample with a
high proportion of GB members), the value of the Oort constant $B$
is much greater in absolute value than expected for the LGD: -13.6
$\pm$ 2.0 $< B <$ -20.7 $\pm$ 1.4 km s$^{-1}$ kpc$^{-1}$. On the
other hand, they find a quite small value of the Oort constant $A$:
10.5 $\pm$ 2.1 $< A <$ 5.7 $\pm$ 1.4 km s$^{-1}$ kpc$^{-1}$. Both
estimations are perfectly compatible with our result of ($A$, $B$) =
(11 $\pm$ 2, -20 $\pm$ 2) km s$^{-1}$ kpc$^{-1}$ for the GB, which
is undoubtedly contaminating their sample.

Some other recent studies find values of $A$ and $B$ in good
agreement with ours. \citet{Uem00}, for a sample of O-B5 {\it
Hipparcos\/} stars, estimate that $A$ = 14.0 $\pm$ 0.7 km s$^{-1}$
kpc$^{-1}$ and $B$ = -15.8 $\pm$ 0.7 km s$^{-1}$ kpc$^{-1}$;
\citet{Zhu00}, working with O-B5 stars with {\it Hipparcos\/} proper
motions, finds that $A$ = 16 $\pm$ 1 km s$^{-1}$ kpc$^{-1}$ and $B$
= -15.6 $\pm$ 0.8 km s$^{-1}$ kpc$^{-1}$; \citet{Oll03}, for a
sample of young main sequence stars find that $(A,B)$ = (9.6, -11.6)
km s$^{-1}$ kpc$^{-1}$, while for a sample of red giants, with no
significant contribution from the GB, the Disk rotation curve is
almost flat, $(A,B)$ = (15.9, -16.9) km s$^{-1}$ kpc$^{-1}$, in very
good agreement with our results for the OB stars of the LGD.

Our estimation of the circular rotation speed, considering that the
Sun's distance to the Galactic center is 8.5 kpc \citep{Ker86}, is
$\Theta_{\odot}$ = 272 km s$^{-1}$ for the LGD from proper motions.
It is a high value, but in agreement with recent studies of O-B5
star samples from the {\it Hipparcos\/} catalog by \citet{Miy98},
$\Theta_{\odot}$ = 268.7 $\pm$ 11.9 km s$^{-1}$; \citet{Uem00},
$\Theta_{\odot}$ = 255.52 $\pm$ 8.33 km s$^{-1}$; or \citet{Bra02},
$\Theta_{\odot}$ = 258.7 $\pm$ 34.29 km s$^{-1}$. A similar result
($\Theta_{\odot}$ = 270 km s$^{-1}$) is also reached by
\citet{Men00} using data from the Southern Proper Motion catalog
\citep{Pla98}; thus, their result is obtained from measures of the
proper motions independent from those of {\it Hipparcos\/}.

\section{Conclusions}

In Paper I we concluded that in the young Galactic disk there are
two spatially different systems, the GB and the LGD. Now we have
classified only by spatial criteria a sample of O-B6 stars from the
{\it Hipparcos\/} catalogue, and found that the GB's global
kinematics is essentially different from the kinematics of the LGD.
Not only that, multidimensional two-samples tests prove that the GB
and the LGD are separated systems in the phase space. This does not
necessarily imply that the GB is a coherent structure born from a
single source such as a great molecular cloud. We just note that in
its present state, the GB is a local system (whose size is well
represented by our sample) showing very different kinematic
properties than the larger system in which it is embedded, the
Galactic disk. The fact that the GB is mainly composed of certain
moving groups challenges the idea of this system coming from a
single origin, and rises the question of whether we are witnessing a
coherent structure with physical entity or we are just observing a
transitory picture of several smaller systems with no common origin
at all. The answer to this question can only be sought in the
dynamical study of the moving groups that form the GB (which would
require a deep knowledge of the Galactic potential and its
asymmetries), in order to trace back their trajectories in the past
and discover whether they come from a single protostellar cloud or
they are the transitory result of some dynamical traps.

We have also proved that the classic problem of the negative vertex
deviation of young stars in the solar neighborhood is caused by the
contribution of the GB. This effect disappears once this stellar
system is removed from the sample, leaving the LGD -defined by its
spatial distribution- as the only remaining structure.

Finally, we have observed how the presence of what is called the GB
introduces disturbances in the estimation of the Oort constants that
describe the kinematics of the young Galactic disk, making it
necessary to discard its contribution by identifying and removing
the GB members. Once the young Disk is pruned by eliminating the GB
members, we estimate a flat rotation curve with a local velocity
very close to the values calculated by several authors in the last
decade for a wide range of ages.

Thus, although we find that the kinematic analysis is not enough to
decipher the origin of the GB, it is indeed fundamental to
characterize the complexity of the young Galactic disk and to better
understand the different moving groups that form the bulk of the GB
stellar component. A comprehensive explanation of the origin of the
GB will require a dynamical analysis of these moving groups.

\acknowledgments

F.E. wants to thank the Departamento de F\'{\i}sica At\'omica,
Molecular y Nuclear of the Universidad de Sevilla for its support
during this work. We would also like to acknowledge the funding from
MCEyD of Spain through grants AYA2004-05395, and
AYA2004-08260-C03-02, and from Consejer\'{\i}a de Educaci\'on y
Ciencia (Junta de Andaluc\'{\i}a) through TIC-101.

\clearpage

\begin{figure}
\epsscale{.60} \plotone{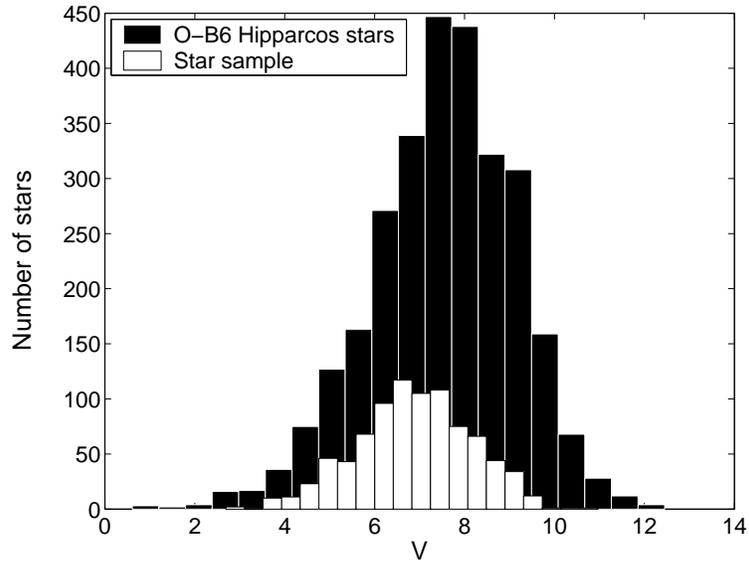} \caption{Histograms of the $V$
magnitude in the Johnson system for the O-B6 stars in the {\it
Hipparcos\/} catalogue \citep{ESA97} and for our star sample.}
\end{figure}

\clearpage

\begin{figure}
\epsscale{.70} \plotone{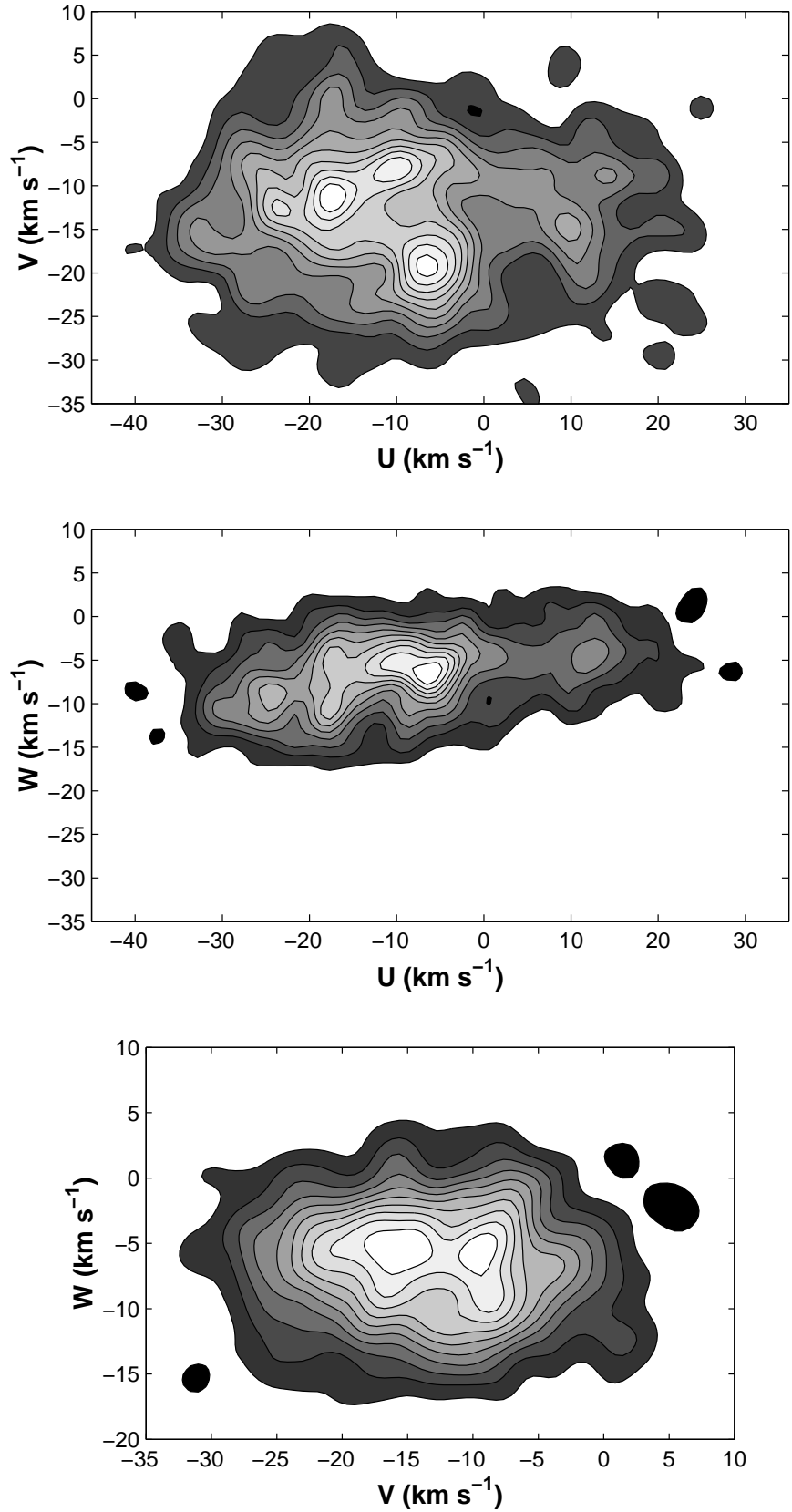} \caption{Distribution of space
velocities for the star sample.}
\end{figure}

\clearpage

\begin{figure}
\epsscale{.80} \plotone{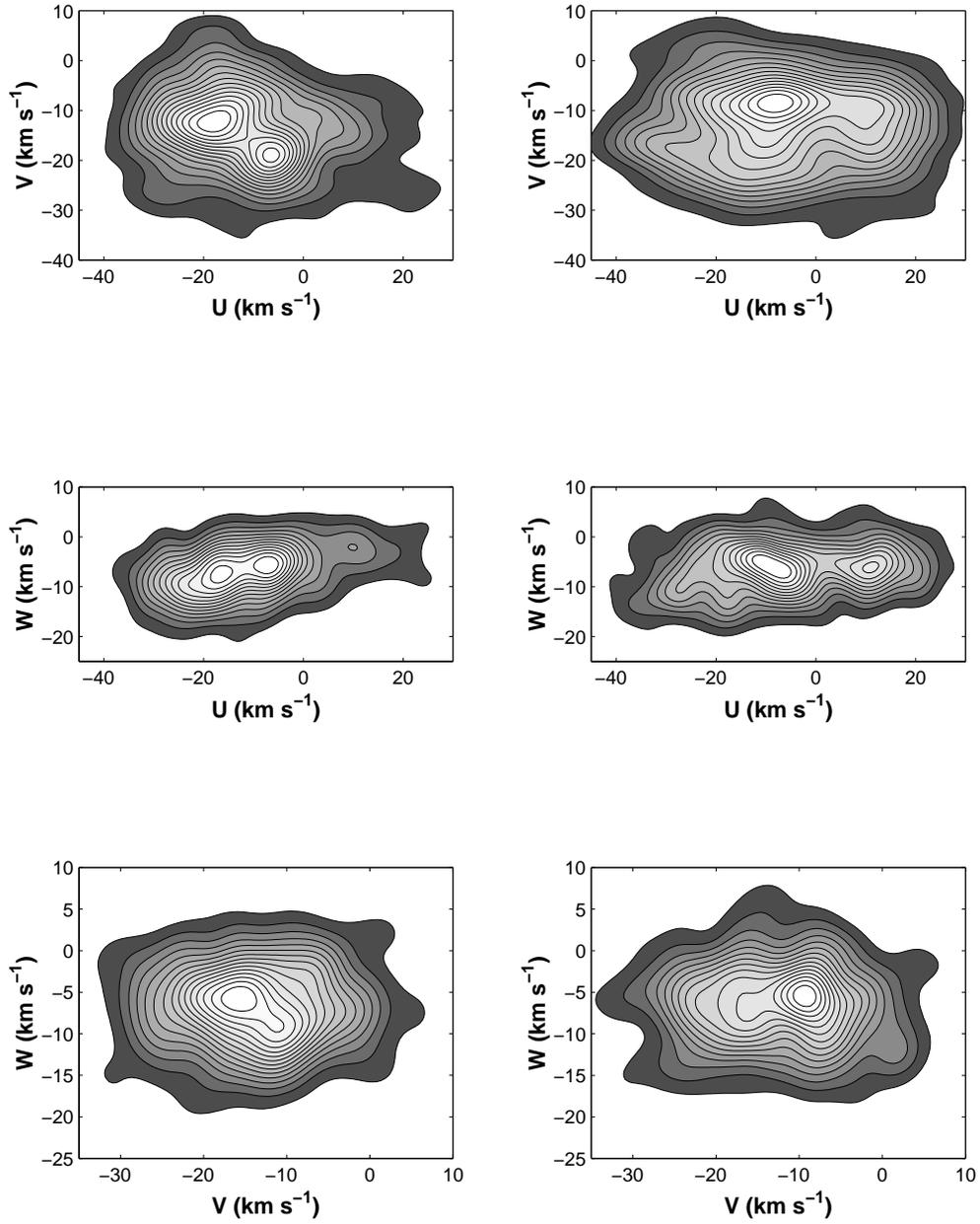} \caption{Distribution of space
velocities for the GB (left) and the LGD (right) stars.}
\end{figure}

\clearpage

\begin{figure}
\epsscale{.80} \plotone{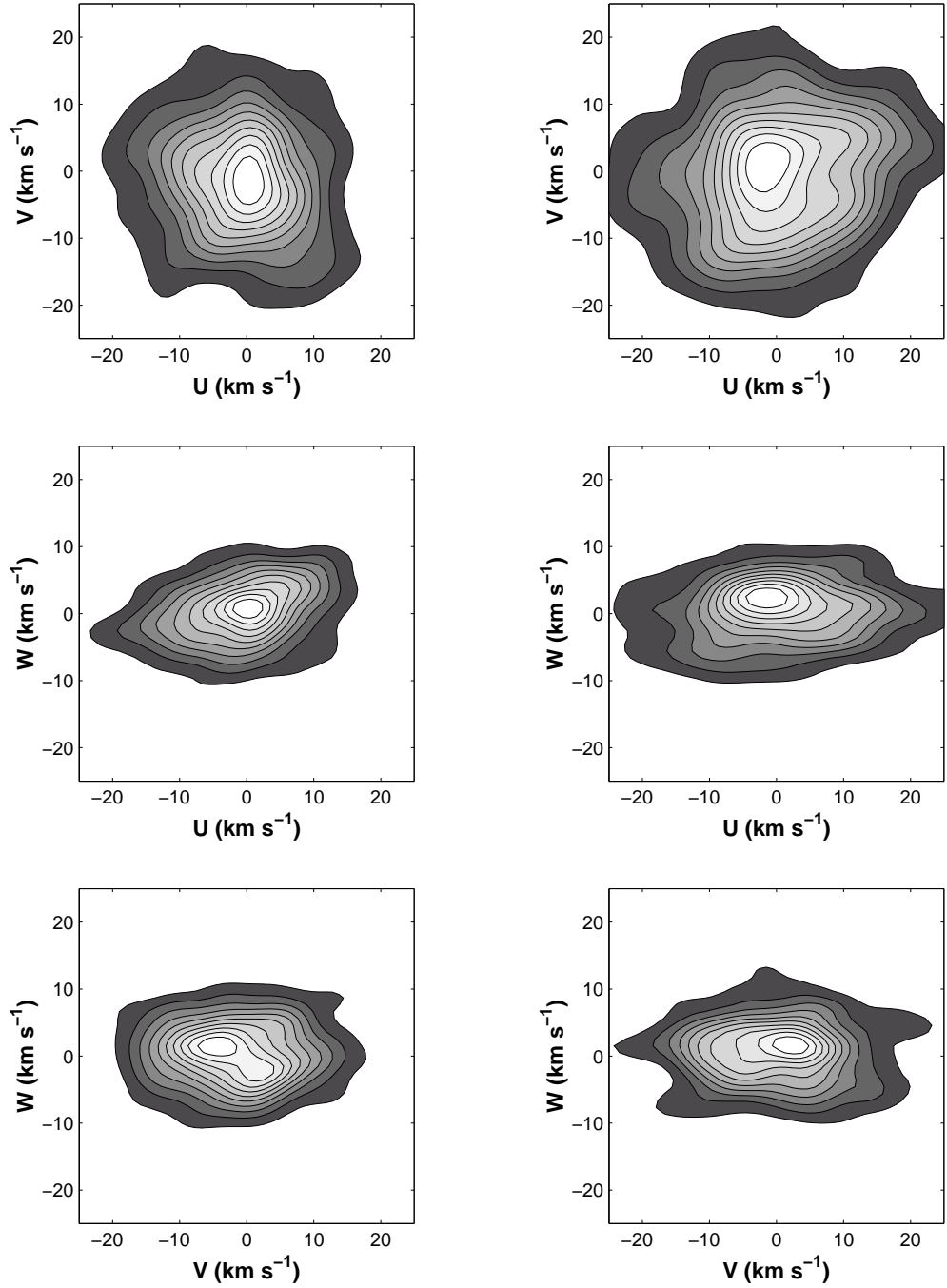} \caption{Contour density plots of
the residual velocities for the GB (left) and the LGD (right)
stars.}
\end{figure}


\clearpage

\begin{thebibliography}{}
\bibitem[Arenou \& Luri(1999)]{Are99} Arenou, F., \& Luri, X. 1999, ASPC, 167, 13
\bibitem[Asiain et al.(1999)]{Asi99} Asiain, R., Figueras, F., Torra, J., Chen, B., 1999, \aap, 341, 427
\bibitem[Baringhaus \& Franz(2004)]{Bah04} Baringhaus, L., Franz, C., 2004, JMVA, 88, 190
\bibitem[Balona \& Shobbrook(1987)]{Bal87} Balona, L. A., Shobbrook, R. R., 1987, \mnras, 211, 375
\bibitem[Barbier-Brossat \& Figon(2000)]{Bar00} Barbier-Brossat, M., Figon, P., 2000, \aap, 142, 217
\bibitem[Blaauw(1956)]{Bla56} Blaauw, A., \apj, 123, 408
\bibitem[Bonneau(1964)]{Bon64} Bonneau, M., 1964, JO, 47, 251
\bibitem[Branham(2002)]{Bra02} Branham, R.L., Jr., 2002, \apj, 570, 190
\bibitem[Brown(2002)]{Bro02} Brown, A.G.A., 2002, ASPC, 285, 150
\bibitem[Cabrera-Ca\~no \& Alfaro(1985)]{Cab85} Cabrera-Ca\~no, J., Alfaro,
E.J., 1985, \aap, 150, 298
\bibitem[Chen et al.(1997)]{Che97} Chen, B., Asiain, R., Figueras F., Torra, J., 1997, \aap, 318, 29
\bibitem[Clube(1972)]{Clu72} Clube, S.V.M., 1972, \mnras, 159, 289
\bibitem[Comer\'on(1992)]{Com92} Comer\'on, F., 1992, PhD Thesis, Universitat de
Barcelona, Spain
\bibitem[Comer\'on(1999)]{Com99} Comer\'on, F., 1999, \aap, 351, 506
\bibitem[Comer\'on \& Torra(1991)]{Com91} Comer\'on, F., Torra, J., 1991, \aap, 241, 57
\bibitem[Comer\'on et al.(1994)]{Com94} Comer\'on, F., Torra, J., G\'{o}mez, A.E., 1994, \aap, 286, 789
\bibitem[de Zeeuw et al.(1999)]{Zeu99} de Zeeuw, P.T., Hoogerwerf, R., de Bruijne, J.H.J., Brown, A.G.A., \& Blaauw, A. 1999, \aj, 117, 354
\bibitem[Dehnen(1998)]{Deh98} Dehnen, W., 1998, \aj, 115, 2384
\bibitem[Delhaye(1965)]{Del65} Delhaye, J., 1965, in Galactic Structure, A. Blaauw \& M. Schmidt, University of Chicago Press, Chicago
\bibitem[Eggen(1963)]{Egg63} Eggen, O.J., 1963, \aj, 68, 697
\bibitem[Eggen(1996)]{Egg96} Eggen, O.J., 1996, \aj, 112, 1595
\bibitem[Elias et al.(2006)]{Eli06} Elias, F., Cabrera-Ca\~no, J.,
Alfaro, E., 2006, \aj, accepted 30 January 2006
\bibitem[Elmegreen(1982)]{Elm82} Elmegreen, B.G., 1982,
in Submillimeter Wave Astronomy, eds. J.E. Beckman \& J.P. Phillips
\bibitem[Elmegreen et al.(2000)]{Elm00} Elmegreen, B.G., Efremov, Y., Pudritz, R.E., Zinnecker, H., 2000, in Protostars and Planets IV,
eds. V. Mannings, A.P. Boss, S.S. Russell, Tucson: University of
Arizona Press
\bibitem[ESA(1997)]{ESA97} ESA, 1997, The {\it Hipparcos\/} and {\it Tycho\/} Catalogues, ESA SP-1200
\bibitem[Fern\'andez(2005)]{Dav05} Fern\'andez, D. 2005, PhD Thesis, Universitat de Barcelona, Spain
\bibitem[Filin(1957)]{Fil57} Filin, A.Ia., 1957, Soviet Astron., 1, 517
\bibitem[Franz(2004)]{Fra04} Franz, C., 2004, cramer: Multivariate nonparametric Cramer-Test
  for the two-sample-problem. R package version 0.7-1.
\bibitem[Fricke \& Tsioumis(1975)]{Fri75} Fricke, W., Tsioumis, A., 1975, \aap, 42, 449
\bibitem[Frogel \& Stothers(1977)]{Fro77} Frogel, J.A., Stothers, R., 1977, \aj, 82, 890
\bibitem[Gould(1879)]{Gou79} Gould, B.A., 1879, Uranometr\'{\i}a
Argentina, Result. Obs. Nac. Argentino, I, p. 354
\bibitem[Grenier et al.(1999)]{Gre99} Grenier, S., Baylac, M.-O., Rolland, L., Burnage, R.,
Arenou, F., Briot, D., Delmas, F., Duflot, M., Genty, V., G\'{o}mez,
A.E., Halbwachs, J.-L., Marouard, M., Oblak, E., Sellier, A., 1999,
\aaps, 137, 451
\bibitem[Guillout et al.(1998)]{Gui98} Guillout, P., Sterzik, M.F., Schmitt, J.H.M.M., Motch, C., Neuh\"auser,
R., 1998, \aap, 337, 113
\bibitem[Hauck \& Mermilliod(1998)]{Hau98} Hauck, B., Mermilliod, M., 1998, \aaps, 129, 431
\bibitem[Herschel(1847)]{Her47} Herschel, J.F.W., 1847, Results of Astron.
Observations made during the years 1834-1838 at the Cape of Good
Hope, London
\bibitem[Johnson \& Soderblom(1987)]{Joh87} Johnson, D.R.H., Soderblom, D.R., 1987, \aj, 93, 864
\bibitem[Kaltcheva \& Knude(1998)]{Kal98} Kaltcheva, N., Knude, J., 1998, \aap, 337, 178
\bibitem[Kerr \& Lynden-Bell(1986)]{Ker86} Kerr, F.J., Lynden-Bell, D., 1986, \mnras, 221, 1023
\bibitem[Lesh(1968)]{Les68} Lesh, R.J., 1968, \apjs, 17, 371
\bibitem[Lindblad(1967)]{Lin67} Lindblad, P.O., 1967, BAN, 19, 34
\bibitem[Ma\'{\i}z-Apell\'aniz(2001)]{Mai01} Ma\'{\i}z-Apell\'aniz, J. 2001, \aj, 121, 2737
\bibitem[Ma\'{\i}z-Apell\'aniz et al.(2004)]{Mai04} Ma\'{\i}z-Apell\'aniz, J., Walborn, N.R., Galu\'e, H.\'A., Wei, L.H., 2004, \apjs, 151, 103
\bibitem[Ma\'{\i}z Apell\'aniz(2005)]{Mai05} Ma\'{\i}z Apell\'aniz, J., 2005, ESA SP-576: The Three-Dimensional Universe
with Gaia, 179
\bibitem[M\'endez et al.(2000)]{Men00} M\'endez, R.A., Platais, I., Girard, T.M., Kozhurina-Platais, V., van Altena, W.F., 2000, \aj, 119, 813
\bibitem[Meynet et al.(1993)]{Mey93} Meynet, G., Mermilliod, J.-C.,
Maeder, A., 1993, \aaps, 98, 477
\bibitem[Montes et al.(2001)]{Mon01} Montes, D., L\'opez-Santiago, J., G\'alvez,
M.C., Fern\'andez-Figueroa, M.J., De Castro, E., Cornide E., 2001,
\mnras, 328, 45
\bibitem[Moreno et al.(1999)]{Mor99} Moreno, E., Alfaro, E.J., Franco, J., 1999, \apj, 522, 276
\bibitem[Mihalas \& Binney(1981)]{Mih81} Mihalas, D., Binney, J., 1981. Galactic astronomy: Structure and kinematics,
W. H. Freeman and Co., San Francisco, CA
\bibitem[Miyamoto \& Zhu(1998)]{Miy98} Miyamoto, M., Zhu, Z., 1998, \aj, 115, 1483
\bibitem[Olano(1982)]{Ola82} Olano, C.A., 1982, \aap, 112, 195
\bibitem[Olling \& Dehnen(2003)]{Oll03} Olling, R.P., Dehnen, W., 2003, \apj, 599, 275
\bibitem[Perrot \& Grenier(2003)]{Per03} Perrot, C. A., Grenier, I.
A., 2003, \aap, 404, 519
\bibitem[Piskunov et al.(2006)]{Pis06} Piskunov, A.E., Kharchenko, N. V., R\"oser,
S., Schilbach, E., Scholz, R.-D., 2006, \aap, 445, 545
\bibitem[Platais et al.(1998)]{Pla98} Platais, I., Girard, T.M., Kozhurina-Platais, V., van Altena, W.F., L\'opez, C.E., M\'endez R.A., Ma, W-Z, Yang, T-G, MacGillivray, H.T., Yentis, D.J., 1998, \aj, 116, 2556
\bibitem[Proctor(1869)]{Pro69} Proctor, R., 1869, Proc. Roy. Soc. London, 18, 169
\bibitem[R Development Core Team(2005)]{R05} R Development Core
Team, 2005, R: A language and environment for  statistical
computing. R Foundation for Statistical Computing, Vienna, Austria.
ISBN 3-900051-07-0, http://www.R-project.org
\bibitem[Schmidt-Kaler(1982)]{Sch82} Schmidt-Kaler, Th., 1982, Landolt-Bornstein,
Numerical Data and Functional Relationships in Science and
Technology, Vol 2, p. 19, ed. Hellwege K.H., Springer-Verlag, Berlin
\bibitem[Schr\"oder et al.(2004)]{Sch04} Schr\"oder, S.E., Kaper, L., Lamers, H.J.G.L.M., Brown,
A.G.A., 2004, \aap, 428, 149
\bibitem[Smart(1968)]{Sma68} Smart, W.M., 1968, Stellar Kinematics, Wiley, New York
\bibitem[Stauffer et al.(1997)]{Sta97} Stauffer, John R., Hartmann, Lee W., Prosser, Charles F.,
Randich, Sofia, Balachandran, Suchitra, Patten, Brian M., Simon,
Theodore, Giampapa, Mark, 1997, \aj, 113, 740
\bibitem[Stauffer et al.(1999)]{Sta99} Stauffer, J.R., Barrado y
Navascu\'es, D., Bouvier, J., et al., 1999, \apj, 527, 219
\bibitem[Torra et al.(2000)]{Tor00} Torra, J., Fern\'{a}ndez, D., Figueras, F., 2000, \aap, 359, 82
\bibitem[Turon et al.(1992)]{Tur92} Turon, C., et al., 1992, The {\it Hipparcos\/} Input Catalogue, ESA SP-1136
\bibitem[Uemura et al.(2000)]{Uem00} Uemura, M., Ohashi, H., Hayakawa, T., Ishida, E., Kato, T., Hirata,
R., 2000, \pasp, 52, 143
\bibitem[Westin(1985)]{Wes85} Westin, T.N.G., 1985, \aaps, 60, 99
\bibitem[Zhu(2000)]{Zhu00} Zhu, Z., 2000, \pasj, 52, 1133

\end{thebibliography}
\end{document}